\documentclass[letterpaper,USenglish]{lipics-v2021} 
\nolinenumbers
\hideLIPIcs
\usepackage{url}
\hypersetup{colorlinks=true,linkcolor=red,urlcolor=blue,citecolor=blue}
\usepackage[iso,american,cleanlook]{isodate}

\usepackage[export]{adjustbox}
\usepackage{fancyhdr}
\makeatletter\let\ps@plain\ps@fancy\makeatother

\usepackage[maxfloats=150]{morefloats}
\newcommand\Invisible[1]{
  \marginpar{\color{white}{\fontsize{.5}{.5}\selectfont #1 }}
}
 
\usepackage{microtype}%if unwanted, comment out or use option "draft"

\usepackage{listings}
\usepackage{graphicx}
\usepackage{caption}
\usepackage{cite}
\usepackage{url}
\usepackage[bottom]{footmisc}

\newcommand{\remove}[1] {}

\newcommand{\code}[1] {\texttt{#1}}

\makeatletter
\lst@Key{countblanklines}{true}[t]%
    {\lstKV@SetIf{#1}\lst@ifcountblanklines}
    
\lst@AddToHook{OnEmptyLine}{%
    \lst@ifnumberblanklines\else%
       \lst@ifcountblanklines\else%
         \advance\c@lstnumber-\@ne\relax%
       \fi%
    \fi}
\makeatother

\lstdefinestyle{numbers}
{numbers=left, stepnumber=1, numberstyle=\tiny, numbersep=10pt}
\lstdefinestyle{nonumbers}
{numbers=none}

\title{Ready When You Are: Efficient Condition Variables via Delegated Condition Evaluation}
\titlerunning{Ready When You Are} 
\author{Dave Dice}{Oracle Labs}{dave.dice@oracle.com}{https://orcid.org/0000-0001-9164-7747}{}
\author{Alex Kogan}{Oracle Labs}{alex.kogan@oracle.com}{https://orcid.org/0000-0002-4419-4340}{}
\authorrunning{D. Dice and A. Kogan}
\Copyright{Oracle and/or its affiliates}

\EventShortTitle{} 
\date{}

\begin{document}

\maketitle

\begin{abstract}
Multi-thread applications commonly utilize condition variables for communication between threads.
Condition variables allow threads to block and wait until a certain condition holds, and
also enable threads to wake up their blocked peers notifying them about a change to the state of shared data.
Quite often such notifications are delivered to all threads, while only a small number of specific threads is interested in it.
This results in so-called futile wakeups, where threads receiving the notification wake up and resume their execution 
only to realize that the condition they are waiting for does not hold and they need to wait again.
Those wakeups cause numerous context switches, increase lock contention and cache pressure, translating 
into lots of wasted computing cycles and energy.

In this work, we propose to delegate conditions on which threads are waiting to the thread sending notifications.
This enables the latter to evaluate the conditions and send the notification(s) only to the relevant thread(s), 
practically eliminating futile wakeups altogether.
Our initial evaluation of this idea shows promising results, achieving 3-4x throughput improvement over legacy condition variables.

 \end{abstract}

\section{Introduction}

\remove{
Outline:

Condition variables are important and frequently used.

In the common usage pattern, broadcast API is used, resulting in wasted work.

Our idea: delegating condition evaluation to the signaler.

Another benefit: simplified programming with condition variables.
}

Condition variables are an important synchronization construct frequently used in multi-thread applications.
They enable threads to wait until a particular condition occurs (e.g., an item is put into a queue) before resuming the execution 
of a critical section under a lock (e.g., removing that item from the queue and processing it).
The API for condition variables typically includes three calls: \code{wait}, \code{signal} and \code{broadcast}.
The first one is used to wait for a condition by blocking (parking) the calling thread and (atomically) releasing the lock it is holding, 
the second is used to notify one of the waiters that the condition it has been waiting for might have changed, 
while the third one sends this notification to all waiting threads.
Upon receiving such a notification, a thread wakes up, acquires the lock which it was holding when calling \code{wait}, and
re-evaluates the condition\footnote{Strictly speaking, reevaluating a condition is not 
a part of the API, however, this is a necessary, even if frequently overlooked, step to avoid the \emph{spurious wakeup} problem~\cite{c++guidelines, posix-manual}.}
(calling \code{wait} again if the latter does not hold).

Naturally, the \code{broadcast} call is most suitable when it is expected that \emph{all} threads waiting for a condition 
would be able to proceed (i.e., their condition would be satisfied) once they are notified. 
For instance, \code{broadcast} is commonly used to implement barriers in multi-phase computation, letting threads to
synchronize at the boundaries of each phase.
However, we find that often enough \code{broadcast}s are used in software even if only a few (or even just one) threads 
are expected to find their condition satisfied.

Consider, for instance, a producer-consumer pipeline, where producers put items into a queue and wait until the 
status of their item is updated by a consumer.
When the latter happens, the consumer would use broadcast to notify all the producers, yet only one of them would find
that the status it is interested in has indeed changed.
(This is a pattern found, for instance, in LogCabin~\cite{logcabin}, a distributed storage system.)
All other producer threads would return to sleep on the condition variable, not before they are scheduled to 
run, acquire the lock, and find out that the condition they are waiting for has not changed.
Those futile wake-ups translate into lots of wasted computing cycles and energy.
Besides, they generate a so-called thundering-herd effect, where all signaled threads unnecessary contend 
on the same lock, degrading the software performance even further.

In this work we propose the idea of delegating the evaluation of the conditions waited on by threads to the signaling thread.
This enables the signaling thread to send the signal only to threads that are ``ready to go'', passing those
whose conditions do not hold and which would immediately return to sleep otherwise.
Thus, delegated condition evaluation allows to reduce dramatically the amount of wasted work associated with futile wake-ups.
In practice, our idea requires only a slight change in the \code{wait} call API -- to supply a parameter that can be
used by a signaler to evaluate the condition of the waiter\footnote{Some languages, such as C++, support \code{wait} APIs 
that include a predicate parameter~\cite{cpp-condvars}, thus requiring no further change to the user code.}.
This parameter can be, e.g., a function pointer or a lambda function in languages that support that.
Since in most cases the conditions are as simple as comparing one or very few variables to 
some expected value(s), transforming legacy code to use the new API is trivial and can be done in a nearly mechanical fashion.

Beyond potential performance benefits, we believe that our idea also simplifies the implementation of and reasoning about concurrent
algorithms that employ condition variables.
We demonstrate that with a bounded queue implementation, which is a textbook example for using condition variables.

\Invisible{IP embodied -- Oracle Invention Disclosure Accession Number :
 ORA200224-US-NP ``Efficient Condition Variables via Delegated Condition Evaluation''} 

\section{Technical Details}
\subsection{The waiter's side}

To support delegated condition evaluation (DCE), we need to extend the \code{wait} API call to include the predicate that the signaling thread
can evaluate before sending the actual signal to the waiter.
The actual mechanism may differ between languages and runtime environments. 
In Figures~\ref{alg:legacy-wait}--\ref{alg:dce-wait} we show how it can be implemented in C, using function pointers.
(C++11 supports the \code{wait\_for} API, which includes a predicate in the form of a lambda function~\cite{cpp-condvars}).

Given that \code{wait\_dce} ``knows'' the condition of the caller,  it can now guarantee that
the condition would hold upon the return from the call.
Thus, DCE eliminates the common source of bugs associated with the use of condition variables, namely not calling the \code{wait} function in 
a loop~\cite{c++guidelines}.
As a result, beyond performance benefits, DCE can potentially simplify concurrent programming with condition variables.

We also note that the legacy code can be gradually transformed to use DCE, with the \code{wait} call overloaded 
(e.g., through the LD\_PRELOAD mechanism in Unix or Linux) to call \code{wait\_dce} with a trivial predicate that always returns \code{true}.

\begin{figure}[t!]
\begin{minipage}{.5\textwidth}
\begin{lstlisting}[language=C, caption=Legacy \code{wait}, label=alg:legacy-wait, escapechar=|,deletendkeywords={next}, commentstyle=\color{blue}]
int foo() {
  ...
  while (flag != 1)
    wait(cond, mutex);
  ...
}
\end{lstlisting}
\end{minipage}%
\begin{minipage}{.5\textwidth}
\begin{lstlisting}[language=C, caption=\code{wait} with DCE, label=alg:dce-wait, escapechar=|,deletendkeywords={next}, commentstyle=\color{blue}]
int check_flag(void *arg) {
  int *flag = (int *)arg;
  return *flag == 1;
}

int foo() {
  ...
  wait_dce(cond, mutex, check_flag, &flag);
  ...
}
\end{lstlisting}
\end{minipage}
\end{figure}
\subsection{The signaler's side}
To support DCE, the signaling thread iterates over predicates associated with each waiting thread, and sends the actual signal only when
the predicate evaluates to \code{true}.
For \code{signal}, the signaling thread stops iterating after the first such predicate is found; for \code{broadcast}, it evaluates all the predicates.
In practice, one may interpose over the existing \code{signal} and \code{broadcast} calls, requiring no further change to the underlying application.
Alternatively, one may keep the existing calls with their standard semantics, and introduce DCE alternatives, e.g., \code{signal\_dce} and \code{broadcast\_dce}.
The latter approach is useful to efficiently support the case where \code{broadcast} is intended to send a signal to every waiting thread (e.g.,
when implementing a synchronization barrier), avoiding the need to iterate over predicates that are all expected to return \code{true}.

Note that in order to avoid race conditions between the signaling thread and unblocked waiters, \code{signal} and \code{broadcast} should
be called only when holding the corresponding lock.
This is a typical case in legacy code that uses condition variables anyway (in fact, the POSIX API advises always doing that 
``if predictable scheduling behavior is required''~\cite{posix-manual}), and thus does not pose a practical concern.

\section{Bounded Queue with One Condition Variable}
Bounded queue is a classic use case for condition variables-based synchronization between producers
trying to insert items into a buffer (queue) with a limited capacity and consumers trying to remove items from that same buffer.
A common approach is to implement the bounded queue using two condition variables~\cite{using-condvars}, signaled 
by producers (consumers) when they put a new element into the queue to notify their counterpart that the queue is no longer empty (full, respectively).
We note that some implementations use \code{broadcast} instead of \code{signal}~\cite{vmaf-condvars}, e.g., to support waits with timeouts.

DCE allows to simplify this implementation and use just one condition variable, as shown in Figure~\ref{alg:bounded-queue}.
Beyond conceptual simplicity, DCE reduces the memory cost of the implementation, which might be substantial for applications using numerous bounded queues.

We note that some implementations of bounded queues with legacy condition variables also use one condition variables (e.g., see~\cite{wiki-condvars}).
Those implementations combine the \emph{empty queue} condition variable and the \emph{full queue} one, which means that 
all threads are signaled (i.e., \code{broadcast} is used) when a new item is inserted into or an old one is removed from the queue.
This is exactly the inefficiency eliminated with DCE. 

\begin{figure}[t!]
\begin{lstlisting}[language=C, caption=Bounded queue with DCE, label=alg:bounded-queue, escapechar=|,deletendkeywords={next}, commentstyle=\color{blue}]
mutex *m;
condvar *cv;

void enq (queue_t *q, val_t data) {
	lock(m);
	wait_dce(m, cv, is_full, q);
	put(q, data);
	signal(cv);
	unlock(m);
}

val_t deq (queue_t *q) {
	lock(m);
	wait_dce(m, cv, is_empty, q);
	val_t data = remove(q);
	signal(cv);
	unlock(m);
	return data;
}

\end{lstlisting}
\end{figure}

\section{Initial Evaluation}
For our initial evaluation, we implemented a microbenchmark emulating a system that uses producer and consumer threads.
The microbenchmark has an array of N-1 padded slots, each for one of the N-1 consumers, representing items for processing.
A producer thread randomly picks one of the slots, checks that it is empty (its value is 0), and if so, writes 1 into that slot and calls \code{broadcast}.
(If the slot is not empty, meaning the consumer has not processed yet the previous item, the producer spins until it becomes empty).
After that, a producer performs some local work (simulated by running a random number generation loops for a random number of iterations), and
picks a new random slot.
A consumer thread waits until its slot has a value different from $0$ (by calling \code{wait}), and processes the item (by writing $0$ into its slot).
At the end, we report the number of iterations performed by the producer thread (which denotes the number of items it has ``produced'').

We mock up DCE using one condition variable per consumer thread, and one auxiliary list structure (called \code{wait\_list}) as following. 
Each waiting (producer) thread inserts a node into the \code{wait\_list} with a predicate (represented as a pointer to a function), 
an argument to the predicate (that the latter needs for evaluation), and a pointer to the thread's condition variable.
After writing into a slot, the producer thread runs over the list, calling predicates in the corresponding nodes, until it finds the first predicate that 
evaluates to \code{true}. 
In that case, it calls \code{signal} with the condition variable stored in that node, notifying the corresponding consumer that its element is ready.
Note that the producer wakes up at most one consumer thread at a time, as intended, instead of all waiting threads as does \code{broadcast} with
legacy condition variables.

We run the benchmark on the Oracle X6-2 machine featuring two Intel Xeon CPU E5-2630 v4 processors with 10 hyperthreaded cores each.
As common in performance benchmarking, we disable the turbo mode to reduce the effect this mode  
has on performance, making it less predictable.
We vary the number of consumer threads from one to $79$, bringing the total number of threads to twice the capacity of the machine.
We run each experiment $7$ times in the same configuration, and report the mean of the measured throughput.

\begin{figure*}[!tp]
\subfloat[][Throughput]{\includegraphics[width=0.5\linewidth]{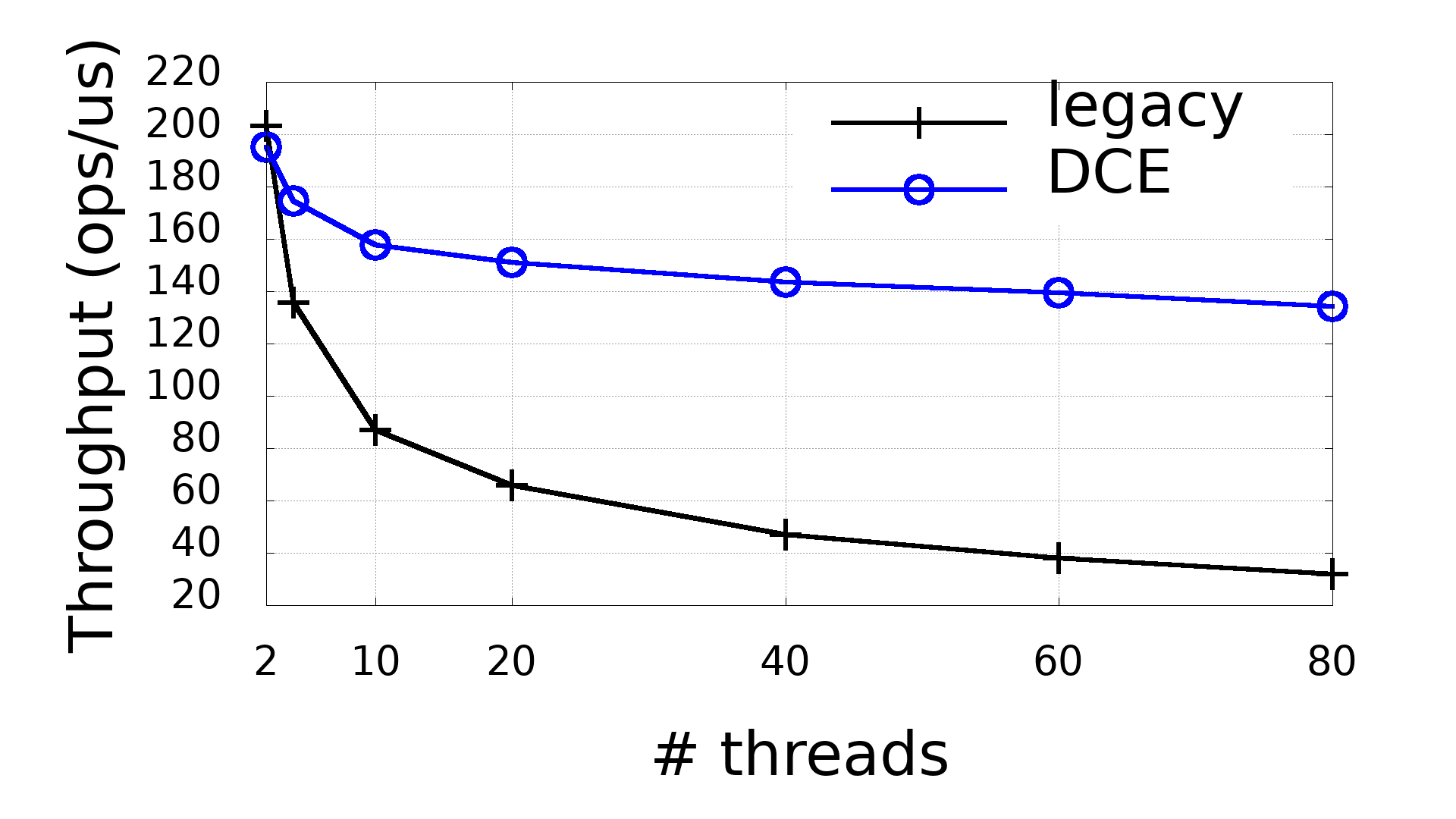}}
\subfloat[][Futile wakeups]{\includegraphics[width=0.5\linewidth]{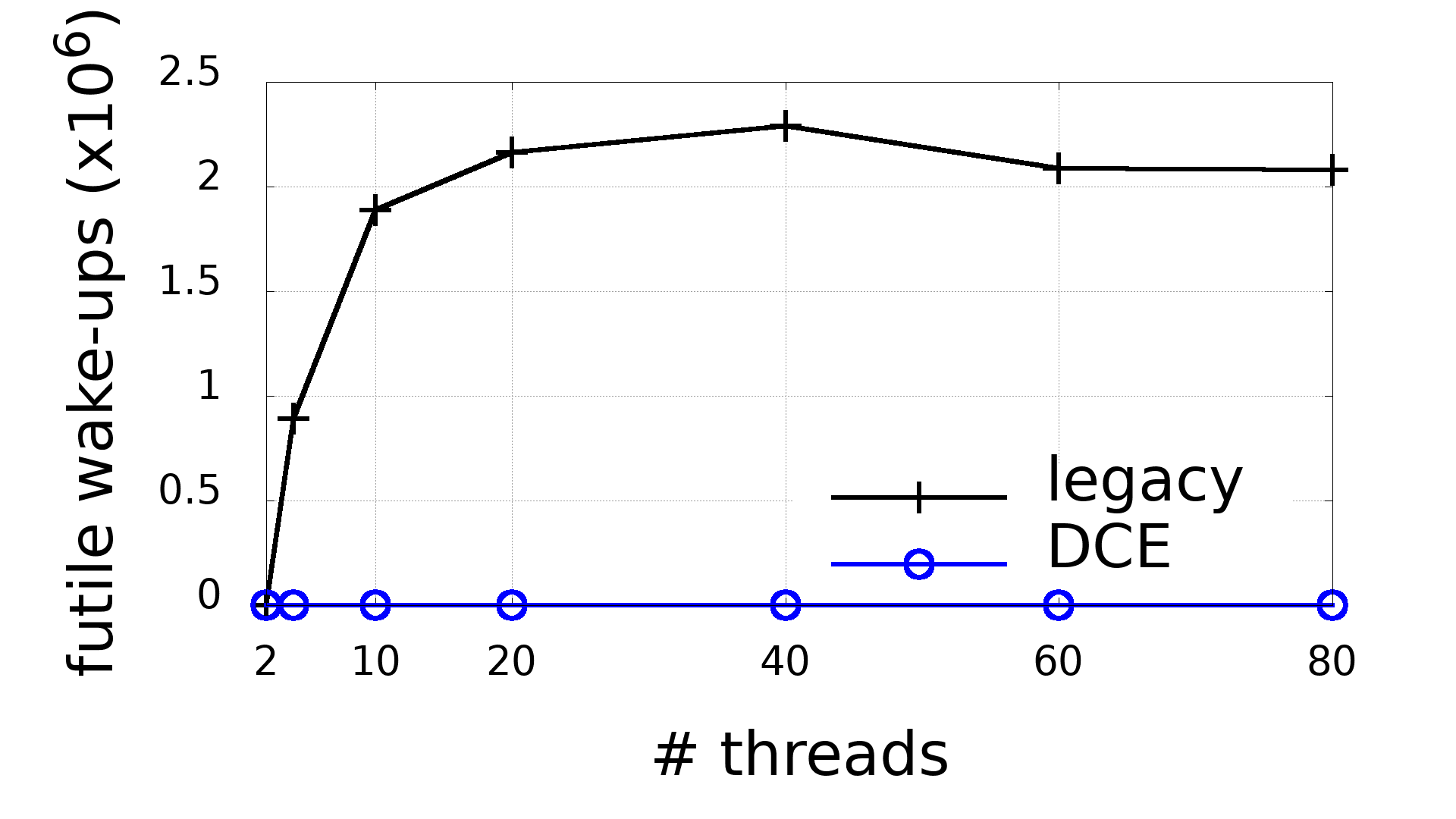}}
\caption{Performance results.}
\label{fig:condvars}
\end{figure*}

The results are presented in Figure~\ref{fig:condvars}~(a).
For a single consumer thread (the first data point on the chart), DCE marginally trails behind the base (legacy) case.
We believe this is because of the small overhead associated with the management of the auxiliary list.
For all other thread counts, DCE substantially outperforms the base case, achieving 3x larger throughput at $40$ threads and 4x at $80$.
Notably, after the initial drop DCE maintains the same level of performance, regardless of the number of consumers.
This is what one would expect from a system with just one consumer.

In the base case, however, throughput fades as we increase the number of consumers.
This is because additional consumers introduce futile wakeups, leading to more context switches and, in general, 
more overhead to process each and every item.
The data in Figure~\ref{fig:condvars}~(b) backs up this hypothesis by showing that in the bace case the number of futile wakeups
grows with the number of consumers (at least, until reaching the capacity of the machine) while DCE introduces no such wakeups.

\section{Extension}
DCE allows a waiting thread to delegate its condition to the signaling thread.
As a possible extension of that idea, the waiting thread may delegate the execution of an action protected by that condition as well.
We call this idea RCV, or remote condition variables, similarly to a related idea of RCL (remote core locking)~\cite{LDT12}, 
in which threads delegate the execution of their critical sections to a dedicated thread.
With RCV a thread passes both the predicate and the action when calling \code{wait}; upon return, its critical section (action)
protected by the predicate is executed.

Unlike RCL that uses dedicated server threads to execute delegated critical sections, in RCV any thread can execute delegated
actions associated with the corresponding condition variable.
Therefore, despite name similarity to RCL, RCV is more closely related to other work on delegated critical section execution, 
e.g.,~\cite{OTY99, KSW18, FK12}.
We note that none of these papers concerns delegation in the context of condition variables.

Note that when \code{wait} returns in RCV, the waiting thread does not hold the lock.
Thus, if it needs to execute a remainder of its critical section, it needs to acquire the lock explicitly.
However, as we discuss next, this allows to reduce the contention on the lock for the cases where the waiter does not need the lock 
beyond the action it has delegated.

Similarly to RCL and other papers cited above, the benefit of RCV comes from reduced contention on the lock and exploiting cache locality.
This is because the signaling thread typically already holds the lock, and thus can execute delegated actions without
lock handoffs.
Those delegated actions are likely to access same cache lines, and thus having the same thread execute the actions 
of all waiting threads (whose conditions are satisfied) reduces the number of cache misses.

We note that DCE can be treated as a degenerate form of RCV, in which only the predicate is delegated.
At the same time, note that delegating actions requires far more invasive changes to the legacy code.
This is because, much like critical sections in RCL  and other cited work, the action code needs to be converted into lambdas or separate functions.
In modern C++ code, however, lambda-based \code{wait} usage becomes widespread, making the switch to RCV more straightforward.

\bibliography{refs}

\end{document}